\documentclass[sigconf]{acmart}

\AtBeginDocument{%
  }

\setcopyright{acmlicensed}

\copyrightyear{2024}
\acmYear{2024}
\setcopyright{rightsretained}
\acmConference[RecSys '24]{18th ACM Conference on Recommender Systems}{October 14--18, 2024}{Bari, Italy}
\acmBooktitle{18th ACM Conference on Recommender Systems (RecSys '24), October 14--18, 2024, Bari, Italy}\acmDOI{10.1145/3640457.3688055}
\acmISBN{979-8-4007-0505-2/24/10}

\usepackage{makecell}  
\usepackage{array}

\begin{document}

\title{Bridging the Gap: Unpacking the Hidden Challenges in Knowledge Distillation for Online Ranking Systems}

\author{Nikhil Khani}
\orcid{0009-0002-8735-8955}
\affiliation{%
  \institution{Google LLC}
  \country{USA}}
\email{nkhani@google.com}

\author{Shuo Yang}
\orcid{0009-0009-1613-4031}
\affiliation{%
  \institution{Google LLC}
  \country{USA}}
\email{yshuo@google.com}

\author{Aniruddh Nath}
\orcid{0009-0006-2815-3035}
\affiliation{%
  \institution{Google LLC}
  \country{USA}}
\email{aniruddhnath@google.com}

\author{Yang Liu}
\orcid{0000-0002-6010-507X}
\affiliation{%
  \institution{Google LLC}
  \country{USA}}
\email{ylyangliu@google.com}

\author{Pendo Abbo}
\orcid{0009-0001-1235-4231}
\affiliation{%
  \institution{Google LLC}
  \country{USA}}
\email{pabbo@google.com}

\author{Li Wei}
\orcid{0009-0008-9321-3983}
\affiliation{%
  \institution{Google LLC}
  \country{USA}}
\email{liwei@google.com}

\author{Shawn Andrews}
\orcid{0009-0004-1129-8196}
\affiliation{%
  \institution{Google LLC}
  \country{USA}}
\email{shawnandrews@google.com}
\author{Maciej Kula}
\orcid{0009-0004-0430-5052}
\affiliation{%
  \institution{Google DeepMind}
  \country{USA}}
\email{maciejkula@google.com}

\author{Jarrod Kahn}
\orcid{0009-0004-5401-3242}
\affiliation{%
  \institution{Google DeepMind}
  \country{USA}}
\email{jarrodk@google.com}
\author{Zhe Zhao}
\orcid{0000-0002-6847-0186}
\affiliation{%
  \institution{University of California, Davis}
  \country{USA}}
\email{zao@ucdavis.edu}

\author{Lichan Hong}
\orcid{0009-0004-9563-554X}
\affiliation{%
  \institution{Google DeepMind}
  \country{USA}}
\email{lichan@google.com}

\author{Ed Chi}
\orcid{0000-0003-3230-5338}
\affiliation{%
  \institution{Google DeepMind}
  \country{USA}}
\email{edchi@google.com}








\renewcommand{\shortauthors}{Khani et al.}

\begin{abstract}
Knowledge Distillation (KD) is a powerful approach for compressing a large model into a smaller, more efficient model, particularly beneficial for latency-sensitive applications like recommender systems. However, current KD research predominantly focuses on Computer Vision (CV) and NLP tasks, overlooking unique data characteristics and challenges inherent to recommender systems.  This paper addresses these overlooked challenges, specifically: (1) mitigating data distribution shifts between teacher and student models, (2) efficiently identifying optimal teacher configurations within time and budgetary constraints, and (3) enabling computationally efficient and rapid sharing of teacher labels to support multiple students. We present a robust KD system developed and rigorously evaluated on multiple large-scale personalized video recommendation systems within Google. Our live experiment results demonstrate significant improvements in student model performance while ensuring consistent and reliable generation of high-quality teacher labels from a continuous data stream of data.
\end{abstract}




\maketitle

\section{Introduction}
Modern recommendation systems demand minimal latency. Even slight delays can negatively impact user experience, especially for large-scale video platforms serving billions of users\footnote{\href {https://www.thinkwithgoogle.com/marketing-strategies/app-and-mobile/page-load-time-statistics/}{https://www.thinkwithgoogle.com/marketing-strategies/app-and-mobile/page-load-time-statistics/}}.  While larger models improve accuracy \cite{cheng2016wide, gomez2015netflix,covington2016deep}, they also increase serving latency, presenting a critical speed and accuracy trade off. Knowledge Distillation (KD) offers a compelling solution by transferring knowledge from a complex "teacher" to a smaller "student" model \cite{hinton2015distilling, cheng2017survey}.  This process allows the student to achieve comparable performance to the teacher without additional latency. The most prevalent method of KD involves three main steps: (1) training a large teacher model on observed data (hard-labels), (2) using teacher to generate predictions on student's training data (soft-labels), (3) training the student on both soft and hard labels.

\noindent
This paper addresses three key challenges in deploying KD to real-world recommender systems: (1) mitigating data distribution shifts between teacher and student models using an online distillation framework with continuous teacher updates and a novel auxiliary task based distillation strategy that allows the student to ground its learning in teacher's knowledge without leaking teacher biases in the student; (2) navigating the costly and time-consuming process of identifying optimal teacher configurations, which can take months (Fig 1), by providing empirically-backed heuristics derived from real-world experiments; and (3) highlighting the overlooked infrastructure challenge of efficiently supporting multiple students distilling from a single teacher for cost amortization and practical deployment.

\section{Challenges \& Distillation Setup}
Our study examines multi-objective pointwise models for ranking videos within a large-scale recommendation system. These models predict short-term objectives like CTR (Click-Through Rate) and long term ones like {\itshape E(LTV)}, estimating the overall value a user drives on the platform over an extended horizon.
\begin{figure}[h]
  \centering
  \captionsetup{justification=centering}
  \includegraphics[width=\linewidth,height=.35\linewidth]{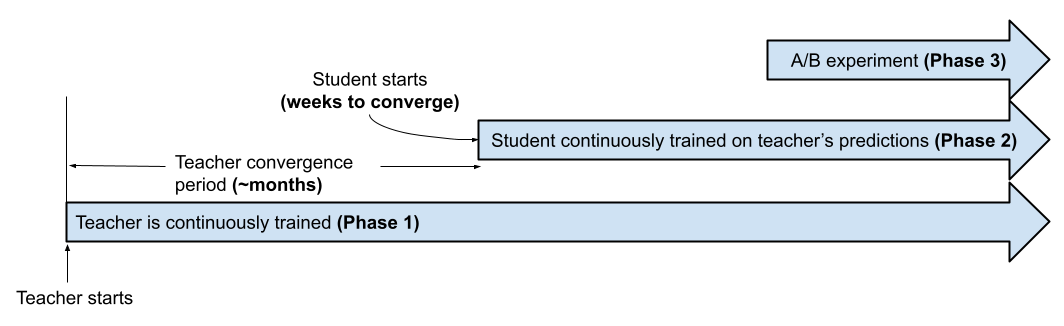}
  \caption{\small\centering A timeline of stages in KD setup}
\end{figure}

\noindent
Our teacher/student models share similar architectures (Fig. 2), beginning with input and embedding layers at the bottom, then multiple shared layers stacked together. The output from the last shared layer feeds into individual towers, producing final per-task prediction logits.\\
\noindent
\\
{\bfseries{\itshape Teacher Setup}} typically utilizes deeper and/or wider shared and task layers compared to the students. To address model divergence common in large models, we employ training stabilization techniques, including learning rate warmup\cite{goyal2017accurate}, activation clipping\cite{krizhevsky2010convolutional}, and Clippy optimizer\cite{tang2023improving}.\\
\noindent
\\
{\bfseries{\itshape Student Setup}} jointly trains on both hard (observed data) and soft labels (teacher predictions) by incorporating an additional distillation loss term. The most commonly used method of distillation, referred as {\itshape“direct distillation”} (Fig 2) uses a single logit to minimize both hard and soft label losses. While effective in CV/NLP with static data distribution, this is suboptimal for recommender systems with rapidly changing data. As an example, we have E(LTV), a noisy and under-calibrated objective. While larger models with more learning capacity can predict LTV accurately, they continue to be under-calibrated  \cite{steck2018calibrated}. Directly using these biased predictions from teachers as guiding examples in students leads to "noise-compounding" \emph{(Challenge \#1)}. This observation highlights that during distillation \textbf{the teacher imparts not only its knowledge, but also its inherent biases to the student}. Our auxiliary distillation strategy addresses this by using separate task logits for hard and soft label losses. This reduces the coupling between observed data and teacher labels, improving the student's ability to leverage teacher knowledge while avoiding bias. Table 1 shows a \textbf{0.4\%} reduction in E(LTV) loss using this technique. Auxiliary distillation has shown to be effective in multiple KD implementations within Google. Additionally, we also continuously train the teacher on new data to ensure up-to-date guidance for students.\\
\noindent
\\
{\bfseries{\itshape Experimentation Time:}} As illustrated in Fig. 1, distillation life-cycle involves months of continuously training teachers and additional weeks of student training and live experiments. This extensive process makes teacher development computationally expensive, especially for new teacher setups \emph{(Challenge \#2)}. To reduce this burden, we offer empirically-backed heuristics based on experimental results (Section 3), addressing questions such as: (1) {\itshape What's an ideal teacher size to train?}, (2) {\itshape Should all objectives be distilled?}, (3) {\itshape If not, which ones to avoid?}

\begin{figure}[h]
  \centering
  \captionsetup{justification=justified}
  \includegraphics[width=\linewidth]{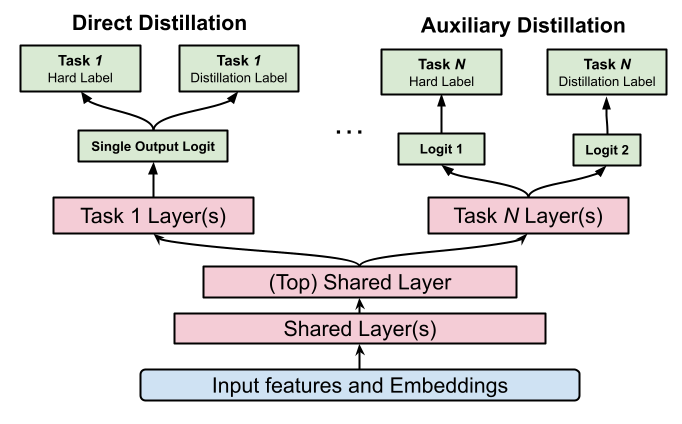}
  \caption{Direct vs Auxiliary distillation}
\end{figure}
\noindent
{\bfseries{\itshape Amortizing Teacher Cost:}} A commonly overlooked infrastructure challenge in KD is the high cost of training and maintaining a teacher. Unlike CV/NLP where data stability allows for less frequent retraining, recommendation systems with dynamic user preferences and item catalogs require continuous model updates. To reduce this burden of continuously updating the large model we amortize its cost across multiple student models served in different contexts (Fig 3) allowing a single teacher model to improve a fleet of students. This requires efficient storage and sharing of teacher labels \emph{(Challenge \#3)}. Our proposed solution involves writing inferences from the trained teacher into a columnar database, prioritizing read performance over strict ACID properties. High data consistency is required to ensure students access the same soft-labels and have similar teacher label coverage. Doing all this on new data (teacher training, label writing, label dissemination) with minimal delay, consistently and reliably is a significant infrastructure challenge. While we use an internal version of BigQuery for our implementation, other columnar databases like Apache Cassandra would work equally well.

\section{Live Experiments}
We evaluate our setup with various teacher configurations and distillation strategies, comparing their offline (AUC for classification, RMSE for regression) and online (engagement, satisfaction) performance against a control model with similar architecture as the student, but trained on hard labels only (no-distillation).\\

\begin{figure}[h]
  \centering
  \captionsetup{justification=centering}
  \includegraphics[height=.55\linewidth]{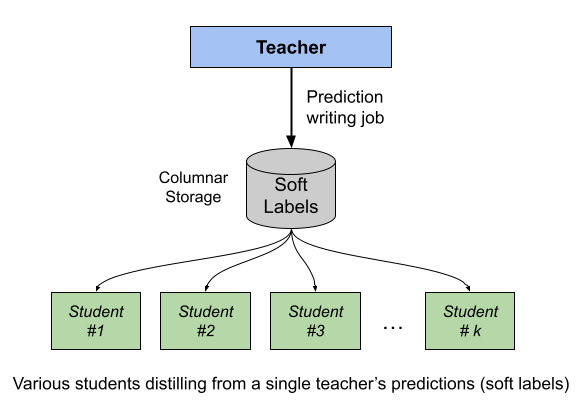}
  \caption{Fleet of students sharing a single teacher's labels}
\end{figure}
\noindent
{\bfseries{\itshape Limiting Bias Leakage in Distillation:}} Table 1 shows the benefit of using auxiliary distillation on noisy tasks, or more generally, as a way to reduce teacher bias from leaking into students. Auxiliary distillation shows a {\bfseries 0.4\%} improvement in E(LTV) compared to direct distillation. In addition, students using direct distillation learn very little from the teacher, as evidenced by the fact that its offline E(LTV) performance is virtually identical to the no-distillation control.

\begin{table}[h]
\centering
\small
\begin{tabular}{l|ccc} 
\toprule
\textbf & \textbf{\makecell{Control \\(no-distillation)}} & \textbf{\makecell{Direct\\Distillation}} & \textbf{\makecell{Auxiliary\\Distillation}} \\ \midrule
E(LTV) RMSE & 12.21 & 12.20 & \textbf{12.15} \\ \bottomrule
\end{tabular}
\caption*{\small Table 1: Effect of distillation strategy on noisy objectives (lower rmse is better).}
\end{table}

\vspace{-6mm}
\noindent
{\bfseries{\itshape Knowledge Distillation By Teacher/Student Size:}} Table 2 shows how student performance improves when distilling even from a relatively small teacher (2x the student size). The 4x teacher yields further gains (additional \textbf{0.43\%} engagement, \textbf{0.46\%} satisfaction). However, this scaling effect of teacher is not expected to continue indefinitely. As demonstrated by \citet{mirzadeh2020improved}, excessively growing the teacher size eventually creates a large knowledge gap between teacher-student making it harder for the student to learn from teacher predictions and hindering the overall performance.

\begin{table}[h]
\centering
\small
\begin{tabular}{c|>{\centering\arraybackslash}m{0.9cm}>{\centering\arraybackslash}m{0.9cm}>{\centering\arraybackslash}m{1.45cm}>{\centering\arraybackslash}m{1.45cm}} 
\toprule
\textbf{\makecell{Teacher\\Size}} & \textbf{\makecell{Teacher\\AUC}} & \textbf{\makecell{Student\\AUC}} & \textbf{\makecell{Engagement\\metric}} & \textbf{\makecell{Satisfaction\\metric}} \\ \midrule
\textbf{\makecell{Control\\(no-distillation)}} & - & 77.86 & - & -  \\ 
\textbf{Large (2x)} & 78.34 & 78.02 & +0.42\% (+0.42\%)* & +0.34\% (+0.34\%)* \\ 
\textbf{X-Large (4x)} & \textbf{78.65} & \textbf{78.06} & \textbf{+0.85\%} (+0.43\%)* & \textbf{+0.80\%} (+0.46\%)*  \\ \bottomrule
\end{tabular}
\caption*{\small Table 2: Student offline (CTR AUC) and online (LE metrics) performance (higher is better), as teacher size increases. All values compared to no-distillation control. Values in parenthesis indicate \% change from the previous row and * denotes statsig with 95\% CI}
\end{table}
\noindent
\\
{\bfseries{\itshape Identifying the Objectives to Distill:}} Determining a priori which tasks benefit from KD in a multi-task  setup is difficult. We investigate this by categorizing model objectives into (1) Primary Engagement Tasks (PET) (2) Primary Satisfaction Tasks (PST) and (3) Others. We train student models with a combinations of these objectives. Table 3 shows that distilling only PET objectives results in the lowest performance gains, aligning with our existing understanding that overemphasizing engagement can promote clickbaity recommendations and harm user satisfaction. Interestingly, distilling both PET and PST objectives led to the highest improvements, even surpassing distilling all objectives (PET+PST+Others). This suggests that while KD targets individual tasks, its effects can spread to shared model layers, potentially causing task conflicts. Thus, a selective approach to task distillation is crucial for optimizing student model performance in multi-task setting.
\begin{table}[h]
\centering
\small
\begin{tabular}{l|cc}
\toprule
Distilled Objectives & \textbf{Engagement Metric} & \textbf{Satisfaction Metric} \\ \midrule
PET & +0.75\% & +0.39\% \\ 
PET + PST & \textbf{+1.13\%} & \textbf{+0.39\%} \\ 
PET + PST + Others & +0.85\% & +0.39\% \\ \bottomrule
\end{tabular}
\caption*{\small\centering Table 3: Effect of distilling various objectives on student quality.}
\end{table}
\vspace{-6mm}
\section{Conclusions And Future Work}
In this paper, we addressed the unique challenges of implementing Knowledge Distillation (KD) in large-scale recommender systems, a domain often neglected in KD research. We introduced an online distillation framework with continuous teacher updates and a novel auxiliary task-based distillation strategy to mitigate data distribution shifts and bias leakage. Additionally, we presented empirically-derived heuristics gleaned from real-world experiments. We recommend starting with a teacher model 2x the size of the student, so it can train and converge faster, without being so over-parameterized as to increase the knowledge gap between teacher and student. And, while the ideal set of distilled objectives is context-dependent, we recommend prioritizing primary engagement and satisfaction tasks, to reduce task conflict during distillation. Removing these tasks from teacher training altogether can further enhance the teacher's accuracy on  PET and PST.\\
Future work in this area will focus on optimizing the latency of teacher label propagation, efficiently training larger model sizes and increasing the breadth of the teacher by including data and objectives from from multiple surfaces while incorporating domain generalization techniques.
\section{Speaker Bio} 
\textbf{Nikhil Khani} is a Senior Software Engineer at YouTube (Google), where he works on improving YouTube's Homepage Ranking.\\
\textbf{Li Wei} is a Senior Staff Engineer at YouTube (Google), working on the WatchNext Team.

\newpage
\bibliographystyle{ACM-Reference-Format}
\bibliography{sample-base}

\appendix

\end{document}